# Hybrid parametric/smooth inversion of electrical resistivity tomography data

Teddi Herring[a]*, Lindsey J Heagy[b], Adam Pidlisecky[a], Edwin Cey[a]

[a] University of Calgary, Department of Geoscience, 2500 University Drive NW, Calgary, Canada, T2N 1N4

[b] University of British Columbia, Department of Earth, Ocean and Atmospheric Sciences, 2329 West Mall, Vancouver, Canada V6T 1Z4

*Corresponding author (tajherri@ucalgary.ca)

**Link to code** https://zenodo.org/record/5116927

**Authorship statement**

Teddi Herring (TH) planned and tested the hybrid inversion strategy with guidance from PhD supervisors Edwin Cey (EC) and Adam Pidlisecky (AP), as well as collaborator Lindsey Heagy (LH). LH provided assistance with navigating and modifying SimPEG geophysical inversion code. TH wrote the manuscript and EC, AP, and LH provided recommendations for improvement.

**Highlights**

- The hybrid inversion approach defines the model space using parametric and smooth components
- Testing scenarios include horizontal surface layers with poorly constrained targets below
- The hybrid algorithm improves resolution of key features for both tested scenarios

# 1    Abstract

The standard smooth electrical resistivity tomography inversion produces an estimate of subsurface conductivity that has blurred boundaries, damped magnitudes, and often contains inversion artifacts. In many problems the expected conductivity structure is well constrained in some parts of the subsurface, but incorporating prior information in the inversion is not a trivial task. In this study we developed an electrical resistivity tomography inversion algorithm that combines parametric and smooth inversion strategies. In regions where the subsurface is well constrained, the model was parameterized with only a few variables, while the rest of the subsurface was parameterized with voxels. We tested this hybrid inversion strategy on two synthetic models that contained a well constrained highly resistive or conductive near-surface horizontal layer and a target beneath. In each testing scenario, the hybrid inversion improved resolution of feature boundaries and magnitudes and had fewer inversion artifacts than the standard smooth inversion. A sensitivity analysis showed that the hybrid inversion successfully recovered subsurface features when a range of regularization parameters, initial models, and data noise levels were tested. The hybrid inversion strategy can potentially be expanded to a range of applications including marine surveys, permafrost/frozen ground studies, urban geophysics, or anywhere that prior information allows part of the model to be constrained with simple geometric shapes.

# 2    Introduction

Electrical resistivity tomography (ERT) is a geophysical method that is used to estimate subsurface electrical resistivity, or its inverse, electrical conductivity. Electrical conductivity depends on lithology, porosity, water saturation, salinity, and temperature. Consequently, ERT has a wide range of applications including geological mapping (Chambers et al., 2012), contaminant imaging (Maurya et al., 2017), geotechnical risk assessment (Perrone et al., 2014), and monitoring dynamic processes like infiltration and drainage (Dietrich et al., 2014) and permafrost thaw (Hilbich et al., 2008).

An ERT survey consists of a series of metal electrodes planted in the earth. Electrical current is injected into the ground between two current electrodes and the resulting voltage difference is measured between two potential electrodes. A series of such measurements is inverted to produce a model of the electrical conductivity of the subsurface, where the model is most commonly parameterized by a large number of voxels. However, the inverse problem is underdetermined and the solution to the inverse problem is nonunique. This issue is usually handled by incorporating regularization that imposes smoothness and/or smallness constraints. The standard ERT inversion produces a model that is a smoothed version of the true conductivity, causing boundaries to be blurred and minimum and maximum values to be damped (Singha and Gorelick, 2005).

Incorporating prior knowledge of the subsurface conductivity in the inversion can mitigate the nonuniqueness problem and yield an estimated model that is consistent with expected subsurface features. How prior information is included in the inversion algorithm depends on the type and reliability of information available (Caterina et al., 2014). In general, there are two approaches to incorporating prior information: (1) modifying the regularization, or (2) modifying the model parameterization. Modifying the regularization incorporates soft constraints. For example, if the subsurface is known to have primarily horizontal bedding, a higher lateral smoothness constraint can be used in the regularization (Oldenburg and Li, 2005). If boundaries between features are expected to be sharp, a L1 model misfit norm may be more appropriate than a L2 norm (Farquharson and Oldenburg, 1998; Loke et al., 2003). Other strategies of modifying regularization include minimizing smoothness across boundary locations (Elwaseif and Slater, 2012), incorporating structure into the reference model (Fortier et al., 2008), including petrophysical and/or geological constraints (Astic and Oldenburg, 2019), and incorporating geostatistical constraints (Hermans et al., 2012). Soft constraints are helpful when prior knowledge is approximate (Kim et al., 2014), but may not sufficiently constrain the precise magnitudes and/or locations of subsurface features (Herring et al. 2021).

When the geometry of subsurface structures is simple and well constrained, prior information can be incorporated using a parametric inversion strategy. Parametric inversion characterizes the subsurface with a small number of model parameters rather than a large number of voxels. In doing so, the inversion problem becomes overdetermined, avoiding the need for regularization. Parametric inversion strategies have been used to describe solute plumes as Gaussian distributions (Pidlisecky et al., 2011), injected tracers as a series of blocks (Lane et al., 2004), and iron bands as elongated Gaussian ellipsoids (McMillan et al., 2016). These studies show that parametric inversion can better recover object boundaries and magnitudes. A caveat to incorporating this type of hard constraint is that if the prior assumptions about model structure are incorrect, the estimated model will be a poor representation of the subsurface.

For many applications of ERT, part of the subsurface is well constrained while the rest is not. Incorporating prior information in these situations to find an optimal conductivity model is challenging. Some studies have incorporated prior information by inverting for a parametric model and using the output as a reference model for a smooth inversion (Goebel, 2019; McMillan et al., 2015). In this study we present a hybrid inversion methodology that combines parametric and smooth model parameterizations and simultaneously solves for both domains. In areas where the model is well-constrained, we use a parametric strategy to characterize the region with only a few parameters, and a smooth inversion fills in the rest of the subsurface conductivity structure. We test this hybrid inversion approach on two different simulated field scenarios with a relatively homogeneous near-surface layer that has a significant conductivity contrast with the underlying earth. Scenario 1 uses ERT to image a conductive contaminant plume beneath a frozen surface layer, and scenario 2 uses ERT to image water infiltration from a surface pond. The goals of this study are to: (1) Present a framework for a hybrid smooth/parametric inversion strategy, (2) test and quantify the success of the hybrid inversion strategy using two simulated field scenarios where the goal is to image a target beneath a well-defined horizontal surface layer, and (3) evaluate the robustness of the hybrid inversion algorithm with respect to regularization, starting model, and data noise.

## 3  Background

### 3.1  *Forward modelling*

An essential step in the inversion process is simulating data for a given conductivity model, called forward modelling. The ERT forward model simulates the electrical potential field ($\boldsymbol{\phi}$) that would be observed for a given conductivity structure ($\boldsymbol{\sigma}$) when a current ($I$) is passed between positive and negative source electrodes at positions $\boldsymbol{r}_{s+}$ and $\boldsymbol{r}_{s-}$, respectively:

$$\nabla \cdot (-\boldsymbol{\sigma}\nabla\boldsymbol{\phi}) = \boldsymbol{I}\big(\delta(\boldsymbol{r} - \boldsymbol{r}_{s+}) - \delta(\boldsymbol{r} - \boldsymbol{r}_{s-})\big) \qquad [1]$$

For most problems, the ERT forward model is solved using finite differences. The model space is discretized so that the electrical conductivity $\boldsymbol{\sigma}$ (S·m$^{-1}$) is characterized as a mesh of conductivity voxels. The electrical potential fields $\boldsymbol{u}$ for a series of source pairs $\boldsymbol{q}$ are a function of electrical conductivity, represented as a diagonal matrix of conductivities $\boldsymbol{S}(\boldsymbol{\sigma})$:

$$(\boldsymbol{DS}(\boldsymbol{\sigma})\boldsymbol{G})\boldsymbol{u} = \boldsymbol{A}(\boldsymbol{\sigma})\boldsymbol{u} = \boldsymbol{q} \qquad [2]$$

where $\boldsymbol{D}$ is the divergence operator and $\boldsymbol{G}$ is the gradient operator. This expression is commonly simplified so that $\boldsymbol{A}(\boldsymbol{\sigma})$ represents the forward operation. A projection matrix $\boldsymbol{P}$ projects the potential field to receiver locations to simulate data $d$:

$$\boldsymbol{d} = \boldsymbol{P}\boldsymbol{A}^{-1}(\boldsymbol{\sigma})\boldsymbol{q} \qquad [3]$$

## 3.2  Smooth inversion

The goal of inversion is to estimate a model that is consistent with the observed data and prior knowledge about the subsurface. An important distinction is the difference between the *model* and the *physical property*. In ERT, the forward operation depends on the physical property $\boldsymbol{\sigma}$, which is defined at each cell in a computational mesh. The model $m$ is what the inversion estimates. A mapping function $\mathcal{M}$ transforms the model into the physical property:

$$\boldsymbol{\sigma} = \mathcal{M}[\boldsymbol{m}] \qquad\qquad [4]$$

In a standard inversion, $\boldsymbol{m}$ and $\boldsymbol{\sigma}$ have a straightforward relationship. They are defined on the same mesh and have the same number of parameters; the only difference is that $\boldsymbol{m}$ is defined as a logarithmic conductivity to enforce a positivity constraint (Loke and Barker, 1996). The mapping function relating the two is an exponential map:

$$\boldsymbol{\sigma} = \mathcal{M}_{exp}[log\,(\boldsymbol{\sigma})] \qquad\qquad [5]$$

Since the standard inversion solves for a model that is defined at each voxel in the mesh, the number of model parameters is larger than the number of data points and the problem is underdetermined, so there are an infinite number of potential solutions. The most common solution is to impose regularization. Tikhonov style inversion minimizes an objective function $\Phi(\boldsymbol{m})$ that contains a data misfit term $\Phi_d$ that penalizes misfit between observed data $\boldsymbol{d}_{obs}$ and predicted data $\boldsymbol{d}(m)$, and a model misfit or regularization term $\Phi_m$ that penalizes differences between the model $\boldsymbol{m}$ and a reference model $\boldsymbol{m}_{ref}$ (Tikhonov and Arsenin, 1977):

$$\Phi(\boldsymbol{m}) = \frac{1}{2}\Phi_d + \frac{\beta}{2}\Phi_m = \frac{1}{2}\|\boldsymbol{W}_d[\boldsymbol{d}(m) - \boldsymbol{d}_{obs}]\|^2 + \frac{\beta}{2}\|\boldsymbol{W}_m(\boldsymbol{m} - \boldsymbol{m}_{ref})\|^2 \qquad [6]$$

$\boldsymbol{W}_d$ is a data weighting term that places lower weight on data with higher measurement error. $\boldsymbol{W}_m$ is a model weighting term that sums a smallness constraint (weighed with a coefficient $\alpha_s$) and directional smoothness constraints (weighted with lateral smoothness coefficient $\alpha_x$ and vertical smoothness coefficient $\alpha_z$ for the 2D problem). $\beta$ is a regularization parameter that determines the relative weight of data misfit and model misfit terms. If $\beta$ is too large the inversion will produce an overly smooth model, and if it too small the inversion may add geologically unreasonable artifacts to the model. There are many approaches to determining a reasonable value of $\beta$ (see Oldenburg and Li (2005)), including fixing the value of $\beta$ at the maximum value that still achieves the desired level of data misfit (Pidlisecky et al., 2007), gradually cooling $\beta$ from a large to small value in successive iterations (Farquharson et al., 2003), the L-curve method (Hansen and O'leary, 1993), or Generalized Cross Validation (Golub et al., 1979).

Linearizing about the reference model and writing the objective function in terms of model update, we get:

$$\Phi(\Delta \boldsymbol{m}) = \frac{1}{2}\|\boldsymbol{W}_d[\boldsymbol{d}(\boldsymbol{m}_{n-1}) + \boldsymbol{J}\Delta\boldsymbol{m} - \boldsymbol{d}_{obs}]\|^2 + \frac{\beta}{2}\left\|\boldsymbol{W}_m\big(\boldsymbol{m}_{n-1} + \Delta\boldsymbol{m} - \boldsymbol{m}_{ref}\big)\right\|^2 \quad [7]$$

where

$$\boldsymbol{J} = \frac{\partial \boldsymbol{d}_{pred}}{\partial \boldsymbol{m}} \qquad [8]$$

is the sensitivity, or Jacobian, matrix that contains the partial derivatives of the predicted data with respect to the model. Rearranging yields a system of equations where we can solve for model update $\Delta \boldsymbol{m}$:

$$(\boldsymbol{J}^T \boldsymbol{W}_d^T \boldsymbol{W}_d \boldsymbol{J} + \beta \boldsymbol{W}_m^T \boldsymbol{W}_m)\Delta \boldsymbol{m} = \qquad [9]$$

$$J^T W_d^T W_d(d(m_{n-1}) - d_{obs}) - \beta W_m^T W_m(m_{n-1} - m_{ref})$$

Computing $J$ directly is computationally expensive. However, in practice we need matrix-vector products of the form $Jv$ and $J^T w$, so $J$ does not need to be explicitly calculated or stored (Haber, 2015). The inexact Gauss-Newton algorithm is a computationally efficient way of calculating an appropriate model update (Pidlisecky et al., 2007). The model is then updated iteratively until a stopping criterion is met.

### 3.3 Parametric inversion

If the geometry of subsurface features is known, it may be appropriate to parameterize the model space with only a few model parameters, rather than at every voxel in the mesh. For example, a two layered model could be described by the conductivity of the first layer, conductivity of the second layer, and depth to the interface. Because of this difference in how the model is defined, there are a few key differences between smooth and parametric inversion. First, the regularization term in the objective function can be discarded in a parametric inversion, as the inverse problem is overdetermined. Second, a mapping function is needed to translate the model parameters to a discretized model of electrical conductivity. Third, a strategy for estimating the sensitivity matrix $J$, or its product with a vector, must be established. The last two points will be discussed below.

Since forward modelling requires that electrical conductivity is defined everywhere on the computational mesh, a parametric mapping function $\mathcal{M}_{par}$ is used to transform the parametric model $m_{par}$ to a conductivity distribution:

$$\boldsymbol{\sigma} = \mathcal{M}_{par}\big[\boldsymbol{m}_{par}\big] \qquad\qquad [10]$$

The nature of the parametric mapping function depends on how the model is parameterized and will be discussed further in the next section.

In a parametric inversion algorithm, there are different ways of calculating the Jacobian. Because there are few model parameters, sensitivities can be calculated with a perturbation approach (Abubakar et al., 2006; Pidlisecky et al., 2011). Another approach, which was used in this study, is to calculate the Jacobian for the full computational model space. Then, using the chain rule and inputting the partial derivatives associated with the mapping function(s), we get the derivative of the measured data with respect to the model parameters (Kang et al., 2015; McMillan, 2017):

$$\boldsymbol{J} = \frac{\partial \boldsymbol{d}_{pred}}{\partial \boldsymbol{m}} = \frac{\partial \boldsymbol{d}_{pred}}{\partial \boldsymbol{\sigma}} \frac{\partial \boldsymbol{\sigma}}{\partial \boldsymbol{m}} \qquad\qquad [11]$$

## 4   Hybrid inversion algorithm

The hybrid inversion algorithm combines strengths of parametric and smooth inversion by dividing the model into two domains and simultaneously updating both. The smooth part is described by a large number of voxels and requires no prior knowledge; the inversion will estimate electrical conductivities for each voxel in whatever way best fits the data while following regularization constraints. The parametric part of the model is described by a small number of parameters to form a simple geometric shape, and the inversion will estimate those parameters with none of the smoothing associated with regularization. In this paper we focus on a specific subset of problems where the survey target is below a horizontally layered feature, so the parametric part of the model is a layer at the surface.

The most unique aspect of the hybrid inversion is the strategic use of mapping functions. The mapping functions make two essential things possible: (1) they contain the instructions for transforming the model into the physical property, and (2) they contain the partial derivatives that are used to calculate the sensitivity matrix with respect to the model parameters. In the examples presented in this study, the parametric model is a simple 2-layer earth, so model parameters that describe the layered model $\boldsymbol{m}_{layer}$ are the logarithmic conductivity of the first layer $log(\sigma_1)$, logarithmic conductivity of the second layer $log(\sigma_2)$, and the depth to the interface $z$:

$$\boldsymbol{m}_{layer} = [log(\sigma_1),\ log(\sigma_2), z] \qquad [1]$$

In this case we assume that the depth to the interface is known, so the "active cells" mapping ($\mathcal{M}_{active\_layer}$) is used to fix $z$. If the layer depth is unknown, $z$ can be left as an inversion parameter. The $\mathcal{M}_{layer}$ mapping function transforms these three values to a 2D discretized model of logarithmic conductivity using a parametric level set approach (Aghasi et al., 2011) to ensure differentiability. Next, the exponential map transforms the model into a 2D conductivity distribution, $\boldsymbol{\sigma}_{par}$ (Figure 1).

$$\boldsymbol{\sigma}_{par} = \mathcal{M}_{exp\_layer}\Big[\mathcal{M}_{layer}\big[\mathcal{M}_{active\_layer}[\boldsymbol{m}_{layer}]\big]\Big] = \mathcal{M}_{par}[\boldsymbol{m}_{layer}] \qquad [2]$$

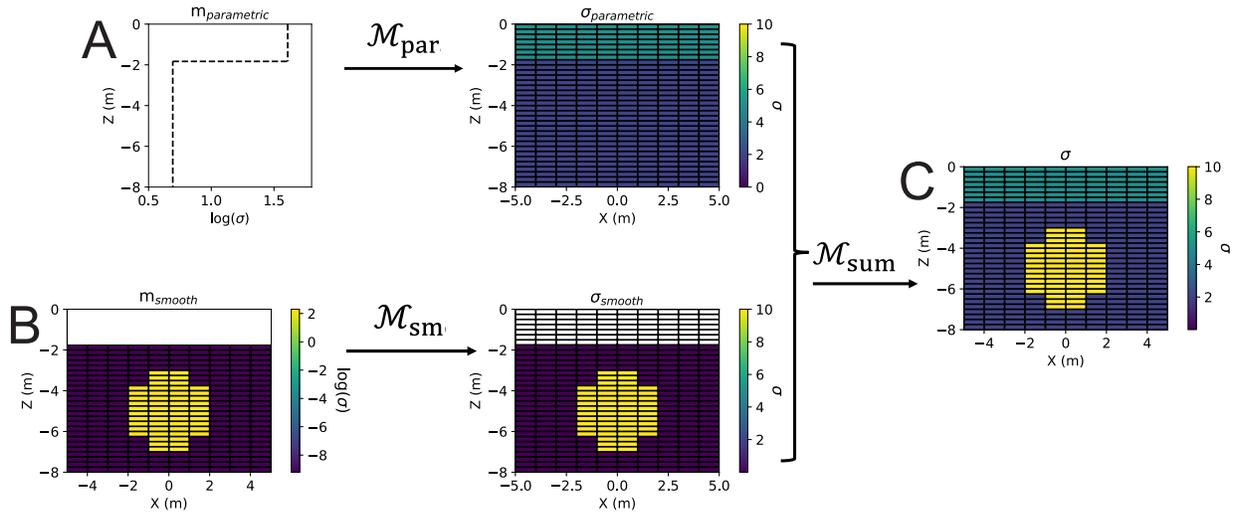

*Figure 1. Summary of the mapping functions. A) The parametric mapping function takes three input values $log(\sigma_1), log(\sigma_2)$, and z, and outputs a discretized 2D of electrical conductivity. B) The smooth mapping function takes a discretized input of logarithmic conductivity in the smooth part of the model space and transforms it to a 2D model of conductivity. C) The summation mapping function adds the parametric and smooth components together to get the full conductivity model.*

For the smooth part of the model, an exponential map transforms $log(\boldsymbol{\sigma})$ into $\boldsymbol{\sigma}$, and then an active cells map $\mathcal{M}_{active\_sm}$ adds cells where the layer will be and sets them to zero:

$$\boldsymbol{\sigma}_{sm} = \mathcal{M}_{active\_sm}\left[\mathcal{M}_{exp\_sm}[\boldsymbol{m}_{sm}]\right] = \mathcal{M}_{sm}[\boldsymbol{m}_{sm}] \qquad [3]$$

To obtain the full discretized electrical conductivity distribution, the smooth and parametric parts are added together with a summation map:

$$\boldsymbol{\sigma} = \mathcal{M}_{sum}[\boldsymbol{m}_{par}, \boldsymbol{m}_{sm}] \qquad [4]$$

Because the parametric and smooth models are summed together, the smooth model is close to zero everywhere but the anomaly location. Solving for a model of electrical conductivity (rather than resistivity) is best so that the model perturbation is small.

Finally, we set up the partial derivatives of each mapping function. The chain rule can be used to calculate the derivative of the electrical conductivity with respect to the model parameters. For the smooth part of the model:

$$\frac{\partial \boldsymbol{\sigma}_{sm}}{\partial \boldsymbol{m}_{sm}} = \frac{\partial \mathcal{M}_{active\_sm}\big[\mathcal{M}_{exp\_sm}[\boldsymbol{m}_{sm}]\big]}{\partial \mathcal{M}_{exp\_sm}[\boldsymbol{m}_{sm}]} \frac{\partial \mathcal{M}_{exp\_sm}[\boldsymbol{m}_{sm}]}{\partial \boldsymbol{m}_{sm}} \qquad [5]$$

The same process can be applied to the parametric part of the model. Once we have these partial derivatives, we can calculate the sensitivity with respect to the desired model parameters (Eq. [11]) and calculate the model update.

This methodology is extensible to other scenarios where the parametric part of the model is a different shape or in a different location in the subsurface. All computational modelling used SimPEG (Cockett et al., 2015), an open source Python-based framework for geophysical simulation and parameter estimation. SimPEG's modular design allows the user to choose parameterizations that best fit their needs, and parametric maps and associated derivatives are already available for a variety of shapes (layers, blocks, spheres, etc.).

# 5   Algorithm testing

In this paper we outline a hybrid inversion strategy where the model space is divided into two domains: a parametric layer at the surface and a smooth model below. To generate synthetic data, we followed the first four steps of the workflow outlined in Figure 2. The last four steps of the workflow describe the hybrid inversion algorithm. We chose to test the algorithm on two synthetic models: one with a target beneath a highly resistive layer at the surface (scenario 1), and one with a target beneath a conductive layer (scenario 2). Scenario 1 represents a saline contaminant plume beneath a frozen surface layer. Scenario 2 represents infiltration from a managed aquifer recharge pond, where the electrodes are located at the base of the pond.

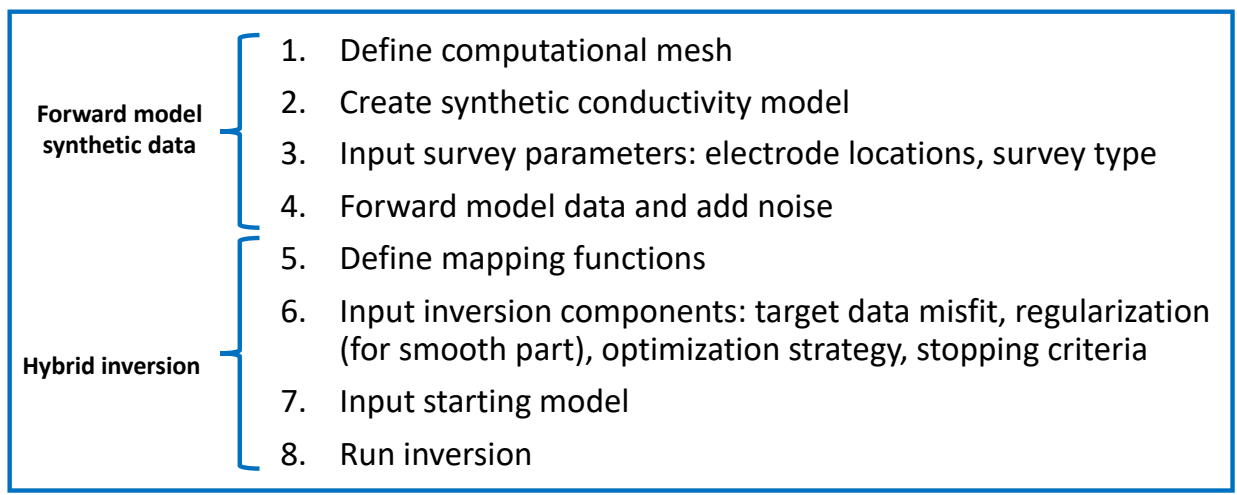

*Figure 2. Workflow to test hybrid inversion algorithm.*

## 5.1   Simulating data

We designed conceptual models using representative site conditions for each of the two scenarios. The conceptual models contain information about geology, pore water saturation and salinity, and temperature. For each conceptual model, the following modified version of Archie's law from Herring et al. (2019) was used to calculate bulk conductivity:

$$\sigma_{bulk} = \begin{cases} a^{-1}\phi^m S_{w0}{}^n k\, C_0[d(T-25)+1] & if\ T \geq 0\ °C \\ a^{-1}\phi^m(S_r S_{w0})^n \dfrac{k\,C_0[d(0-25)+1]S_r}{S_r} & if\ T < 0\ °C \end{cases}$$

[1]

where $\phi$ is porosity (m$^3\cdot$m$^{-3}$), $S_{w0}$ is the initial water saturation (m$^3\cdot$m$^{-3}$), $S_r$ is the relative saturation relating the unfrozen water saturation to the initial saturation, $k$ is a correlation factor between total dissolved solids (TDS) concentration and groundwater conductivity ([S$\cdot$m$^{-1}$] [g$\cdot$L$^{-1}$]$^{-1}$), $C_0$ (g$\cdot$L$^{-1}$) is the TDS concentration of the unfrozen groundwater, $d$ (°C$^{-1}$) relates temperature and fluid conductivity, $T$ is temperature (°C), and $a$, $m$, and $n$ are unitless Archie fitting parameters. The values used for these parameters are listed in Table 1. Relative saturation $S_r$ was determined by the soil freezing characteristic curve from Herring et al. (2019).

*Table 1. Parameters used to calculate bulk conductivity in synthetic modelling examples.*

| | |
|---|---|
| $S_{w0}$ (m$^3\cdot$m$^{-3}$) | Scenario 1: 1.0 Scenario 2: 0.20 |
| $\phi$ (m$^3\cdot$m$^{-3}$) | 0.40 |
| $C_{0\_background}$ (g$\cdot$L$^{-1}$) | 0.50 |
| $a$ (unitless) | 0.36 |
| $m$ (unitless) | 1.3 |
| $n$ (unitless) | 2.3 |
| $k$ ([S$\cdot$m$^{-1}$][g$\cdot$L$^{-1}$]$^{-1}$) | 0.16 |
| $d$ (°C$^{-1}$) | 0.022 |

In each scenario, the ERT survey was designed with 48 electrodes. Each scenario used a dipole-dipole array with the spacing between current and potential dipoles ranging from 1 to 10 times the interelectrode spacing, yielding a total of 408 data points. Since the scale of the two scenarios is different, the electrode spacing was 5 m in the first scenario and 0.5 m in the second scenario. In scenario 1 the smallest cell sizes were 1.0 m in the $x$-direction and 0.25 m in the $z$-direction, while in scenario 2 the smallest cells were 0.125 m in the $x$-direction and 0.0625 m in the $z$-direction. In each case, cell sizes gradually increased towards the edges of the model domain. The total number of cells were 27,560 and 21,700 for scenario 1 and scenario 2, respectively. Forward simulations were computed as a 2.5 dimensional problem using SimPEG (Cockett et al., 2015). When the computational meshes were tested with a homogeneous model, the simulated apparent resistivities had less than 1% error, indicating a reasonable discretization and sufficient model padding.

Data $\boldsymbol{d}_{obs}$ were simulated by forward modelling to get noise-free data $d_{clean}$ and adding noise:

$$\boldsymbol{d}_{obs} = \boldsymbol{d}_{clean} + \varepsilon_{rel} \cdot |\boldsymbol{d}_{clean}| \cdot \boldsymbol{R} \qquad [2]$$

where $\varepsilon_{rel}$ is the relative error and $\boldsymbol{R}$ is a vector of normally distributed random noise, and the standard deviation of the data is equal to $\varepsilon_{rel} \cdot |\boldsymbol{d}_{clean}|$. Initial simulations used a relative data error of 1%, and higher error levels are tested in the sensitivity analysis section below.

## 5.2   Data processing

The inversion algorithm used the mapping functions described in the previous section to create and combine the parametric and smooth parts of the model, using the same computational mesh as the forward model. The smooth part of the model was regularized using a Tikhonov-style approach described previously, with $\alpha_x=\alpha_z=1$, $\alpha_s=10^{-6}$, and $\beta=1$. Sensitivity to $\beta$ is also explored in the sensitivity analysis section below. For both scenarios, the starting background model was set to the average apparent conductivity. The initial layer conductivity was set to the minimum apparent conductivity in scenario 1, and the maximum apparent conductivity in scenario 2. A homogenous starting model for the smooth part of the inversion was assigned a small value of $10^{-4}$ S·m$^{-1}$, as the magnitude of the background is already estimated in the parametric part of the inversion. The data uncertainty was assigned a value of 1%, i.e., the relative data error that was added to the forward modelled data, with a noise floor of 0. In a field experiment the data uncertainty can be estimated using reciprocal measurements. An inexact Gauss-Newton algorithm was used to calculate model updates. The hybrid inversion was allowed to iterate a maximum of 20 steps, or until convergence criteria were met. It is worth noting that other modifications to the smooth inversion discussed above (e.g., other model and data misfit norms, regularization weighting, structured reference models, etc.) can easily be incorporated in the smooth part of the model, but in this case, we assume minimal prior knowledge for the smooth part of the model space.

For comparison, we also processed both synthetic data sets with a standard smooth inversion strategy using the same meshes and survey parameters as the hybrid inversions. We chose to use $\beta=1$, $\alpha_x=\alpha_z=1$, and $\alpha_s=10^{-6}$ in both cases to be consistent with the values used in the hybrid inversion. A homogeneous reference model of the average apparent conductivity was used for the standard inversion. The maximum number of iterations for the smooth inversion was set to 20.

### *5.3 Scenario 1: Contaminant plume in partially frozen ground*

### 5.3.1 Conceptual model

In this synthetic modelling scenario, we created a subsurface conductivity model that included a contaminant plume of leaked fracking flowback water below a frozen surface layer. In this model the plume had a TDS concentration of 150 g·L$^{-1}$ and the background TDS was 0.5 g·L$^{-1}$, which are representative of fracking flowback water and relatively fresh groundwater, respectively (Government of Canada, 2009; Gregory et al., 2011). The plume had a radius of 7 m and its center of mass was at a depth of 15 m (Figure 3A). Subsurface temperature (Figure 3B) was generated with a 1D conductive heat transport model using climate normal data for air temperature and snow depth in northern British Columbia, where most Canadian fracking operations take place (Rivard et al., 2014). Estimates of subsurface thermal conductivity for snow, frozen soil, and unfrozen soil were taken from literature (Herring et al. 2021). The April temperature profile, when the frozen layer reached its maximum thickness of 1.8 m, was used for this conceptual model.

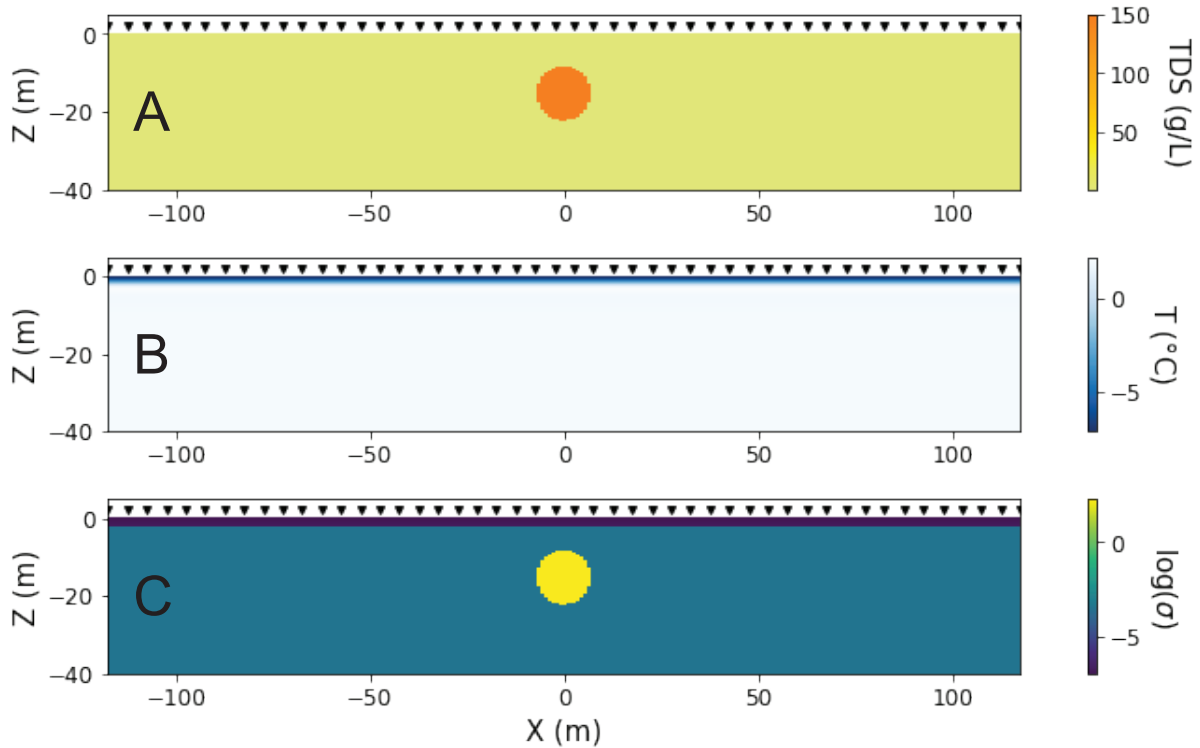

*Figure 3. The key inputs to create a synthetic conductivity model for scenario 1 are A) the total dissolved solids in the groundwater and B) subsurface temperature. Using the petrophysical relationship in Eq. [1] produces the bulk conductivity model (C), which is shown on a logarithmic scale. The black symbols at the ground surface represent electrodes.*

### 5.3.2 Hybrid inversion results

After 20 iterations, the hybrid inversion produced a model with lower data misfit than the standard smooth inversion (Figure 4A). While the standard inversion converged smoothly (Figure 4B), the hybrid inversion converged in a more stepwise manner as different parts of the model updated. In the first seven iterations most of the model updates occurred in the parametric part of the model, and the first large decrease in data misfit occurred as the plume was introduced into the smooth part of the model (Figure 4C). The second jump in data misfit corresponded with a slight decrease in the conductivity of the top layer. The results of the hybrid inversion (Figure 5C) showed a large improvement in the estimated conductivity structure compared to the standard smooth inversion (Figure 5B). Within the frozen layer, the average conductivity was overestimated by 1150% in the standard inversion, whereas the layer conductivity in the hybrid inversion was overestimated by only 20% in the hybrid inversion. The hybrid inversion also recovered a well-defined conductive plume, compared to the standard inversion where the plume was very diffuse. The improved accuracy of the hybrid inversion can also be seen in Figure 6, where vertical profiles of conductivity at locations of $x$=0 m and $x$=75 m were plotted for comparison. The hybrid inversion improved recovery of magnitudes and sharp interfaces, particularly in the top ~7 m. The maximum conductivity in the standard inversion was at 29 m (14 m deeper than the center of mass of the plume), but the hybrid inversion produced a conductivity peak only 0.6 m below the true center of mass. Additionally, the hybrid inversion produced a model without inversion artifacts, which made it easier to interpret the key features. Overall, the hybrid inversion algorithm was a significant improvement from the standard inversion in this case due to the improved resolution of the magnitude and structure of the frozen layer, the more accurate location of the plume, and better overall interpretability.

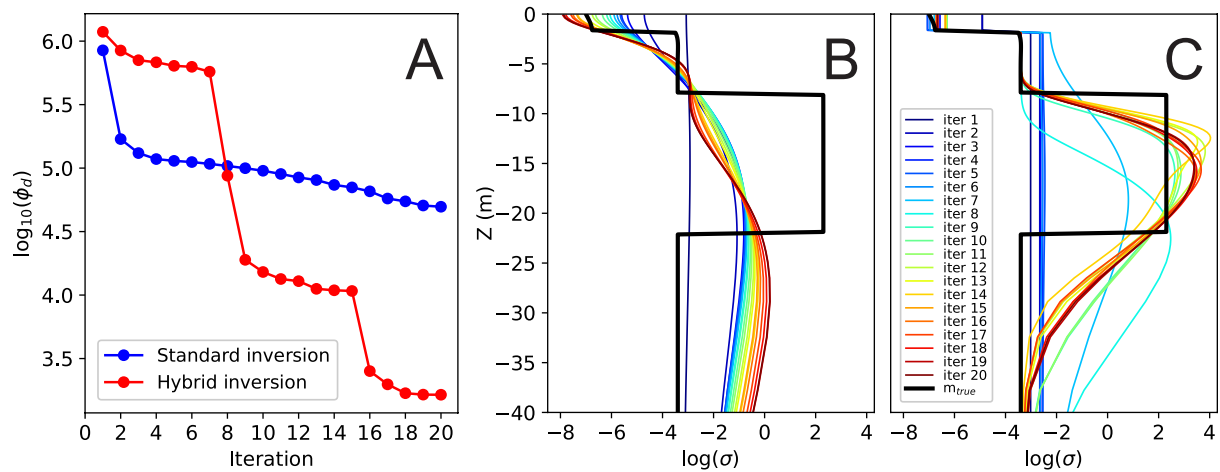

*Figure 4. A) Convergence of the data misfit for standard and hybrid inversions in scenario 1 and vertical slices through models at x=0 m at each iteration of the B) standard and C) hybrid inversions.*

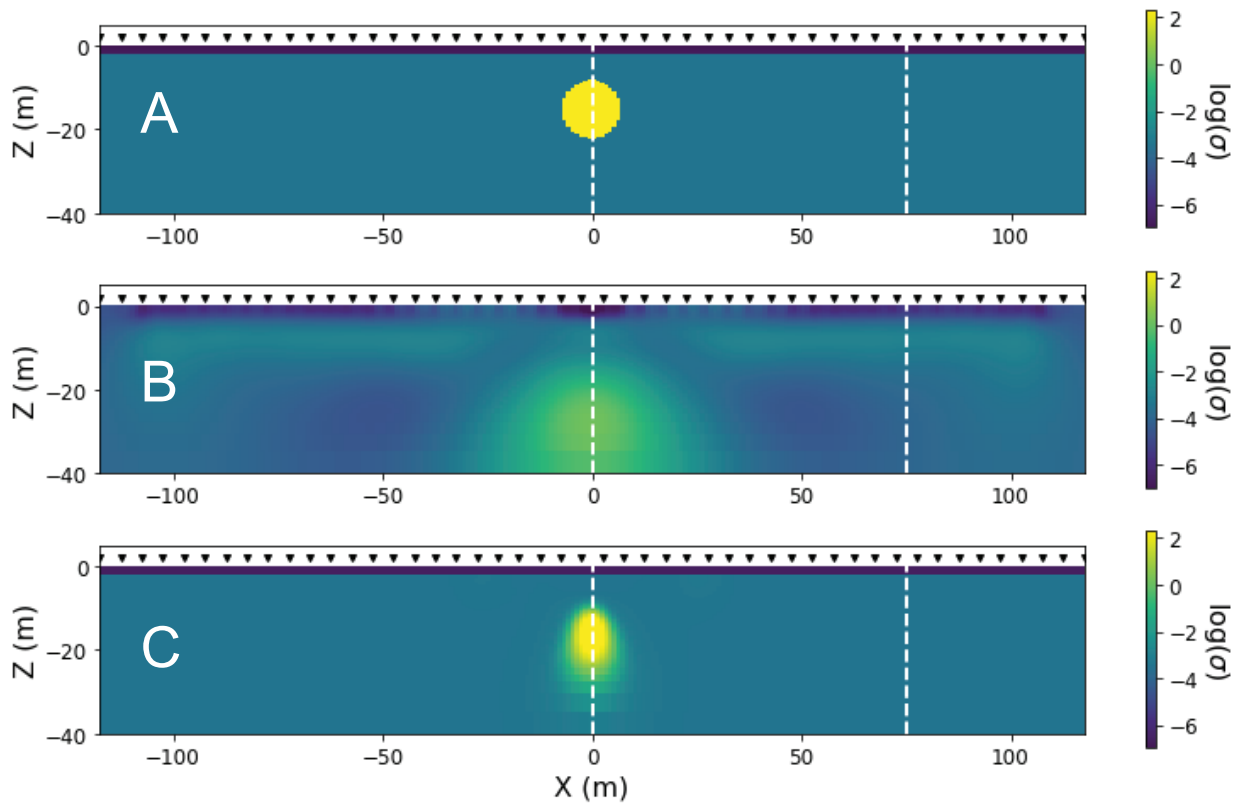

*Figure 5. Scenario 1 inversion results. A) true conductivity distribution, B) estimated conductivity model from a standard inversion, and C) estimated conductivity from a hybrid inversion. The standard inversion produced data with a 10% mean absolute percent error, while the hybrid inversion achieved a mean absolute data error of 2.1%. White dotted lines are the locations of the vertical profiles shown in Figure 6.*

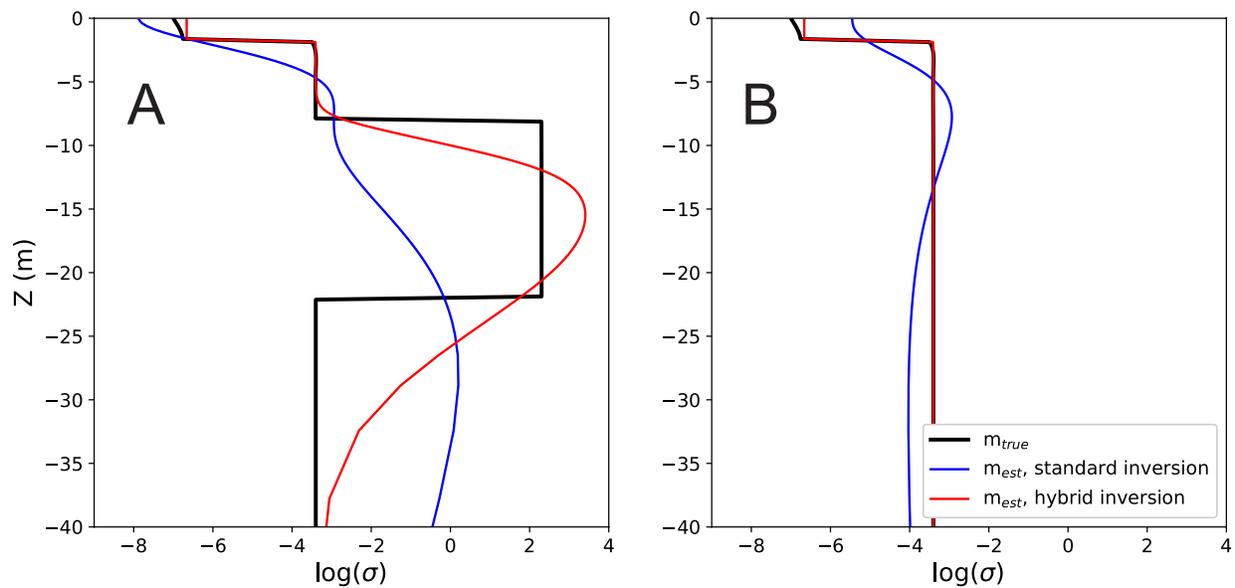

*Figure 6. Vertical profiles of the conductivity models shown in Figure 5 at locations of A) x=0 m and B) x=75 m.*

## 5.4   Scenario 2: Infiltration beneath a recharge pond

### 5.4.1  Conceptual model

In this second hypothetical testing scenario, we simulated an ERT survey that monitored infiltration from a managed aquifer recharge pond, based on a real recharge pond located near Watsonville, California (Haines et al., 2009; Nenna et al., 2011). At this site the pond is filled with water in the winter, and the infiltrated water recharges the aquifer below to meet increased groundwater demand in the dry summer months. The electrodes were installed at the base of the pond and are submerged in water during infiltration.

It is worth noting that in other marine ERT studies the water layer is often defined as a homogeneous layer with a fixed conductivity (Loke and Lane, 2004; Toran et al., 2010), but errors in the water conductivity can dramatically affect the accuracy of the estimated model (Simyrdanis et al., 2015; Tassis et al., 2013). As the hybrid inversion assigns the best-fit water conductivity value, we circumvent the issues associated with inputting a fixed value. Another approach that does not require inputting a fixed conductivity for the water was described by Rücker (2011). Their approach isolates regions of the model space and allows the user to choose the regularization constraints within and across known boundary locations or force a constant property value in a region (Rücker et al., 2017). For a pond scenario similar to the one presented here, their method effectively resolved the conductivity contrast between the water layer and underlying sediments (Rücker, 2011). A drawback of this method is that it requires the user to input boundary locations, which may not always be known.  Although not explored in this paper, the hybrid inversion can also solve for the layer depth if it is unknown. The most appropriate inversion strategy therefore depends on the type of prior knowledge and the associated uncertainty.

As with the previous scenario, we generated the synthetic electrical conductivity model using information about the site. Air temperature from climate normals (US Climate Data, 2020) and estimates of thermal properties (Lal and Shukla, 2004) were used to estimate subsurface temperature with a 1D conductive transport model. The temperature profile for January was used, representing the start of infiltration (Figure 7B). The background saturation was set to $S_w$=0.2, a representative residual saturation for the fines-rich sand at the site (USDA, 2019), and a 0.5 m deep water layer was added above. A saturation profile was estimated using the SimPEG implementation of the Richards equation (Cockett et al., 2018) using fitting parameters from Celia et al. (1990), where the infiltration occurs in a 5m wide region (Figure 7A). Although oversimplified, we believe this is a reasonable starting point from an algorithm testing perspective. Inputting the relevant parameters into Eq. [1] produced the bulk electrical conductivity model in Figure 7C.

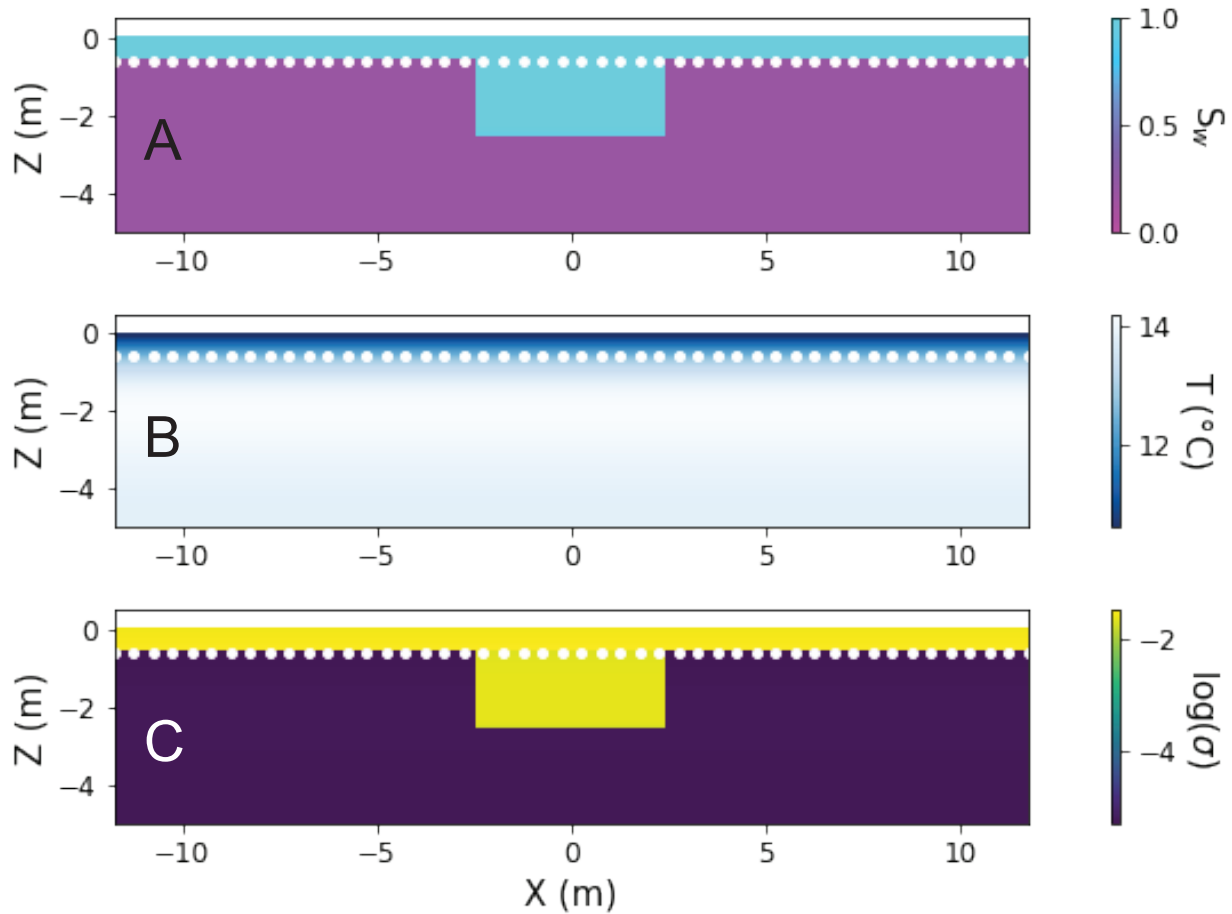

*Figure 7. The key inputs to create a synthetic conductivity model for scenario 2 are A) water saturation and B) subsurface temperature. Using the petrophysical relationship in Eq. [1] produces the bulk conductivity model (C), which is shown on a logarithmic scale. The white symbols below the water layer represent electrodes.*

### 5.4.2 Hybrid inversion results

Since the electrodes in this scenario were located below the water layer, there was no directional information to constrain features above or below the electrodes. This lack of directional information introduces additional ambiguity when searching for an appropriate model update. To improve model convergence, we modified the inversion algorithm to alternate between parametric and smooth model updates. First, the algorithm updated the parametric part of the model with the hybrid inversion method, then updated the smooth part of the model space separately. With this two-step process the data misfit reached a value that was lower than the standard smooth inversion (Figure 8A). The modified hybrid approach produced a good estimated model with minimal inversion artifacts (Figure 9C). In comparison, the standard smooth inversion (Figure 9B) produced a model where the conductive water layer was blurred above and below the electrodes and inversion artifacts made the model difficult to interpret. In this scenario, incorporating prior information with the hybrid inversion resulted in a better estimate of the location and magnitude of the water layer, minimal inversion artifacts, and a more easily interpretable model.

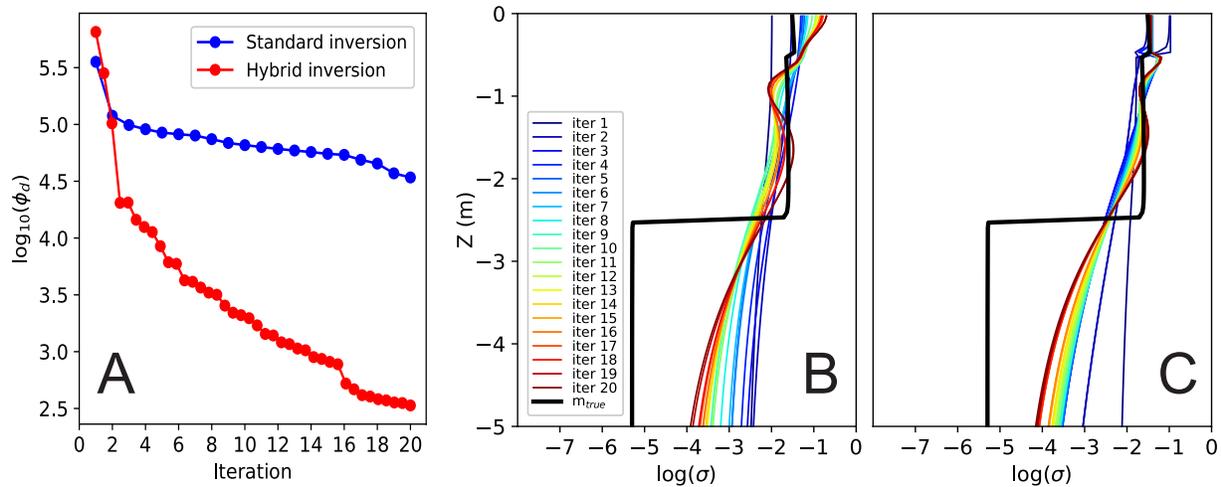

*Figure 8. A) Convergence of the data misfit for standard and hybrid inversions in scenario 2 and vertical slices through models at x=0 m at each iteration of the B) standard and C) hybrid inversions.*

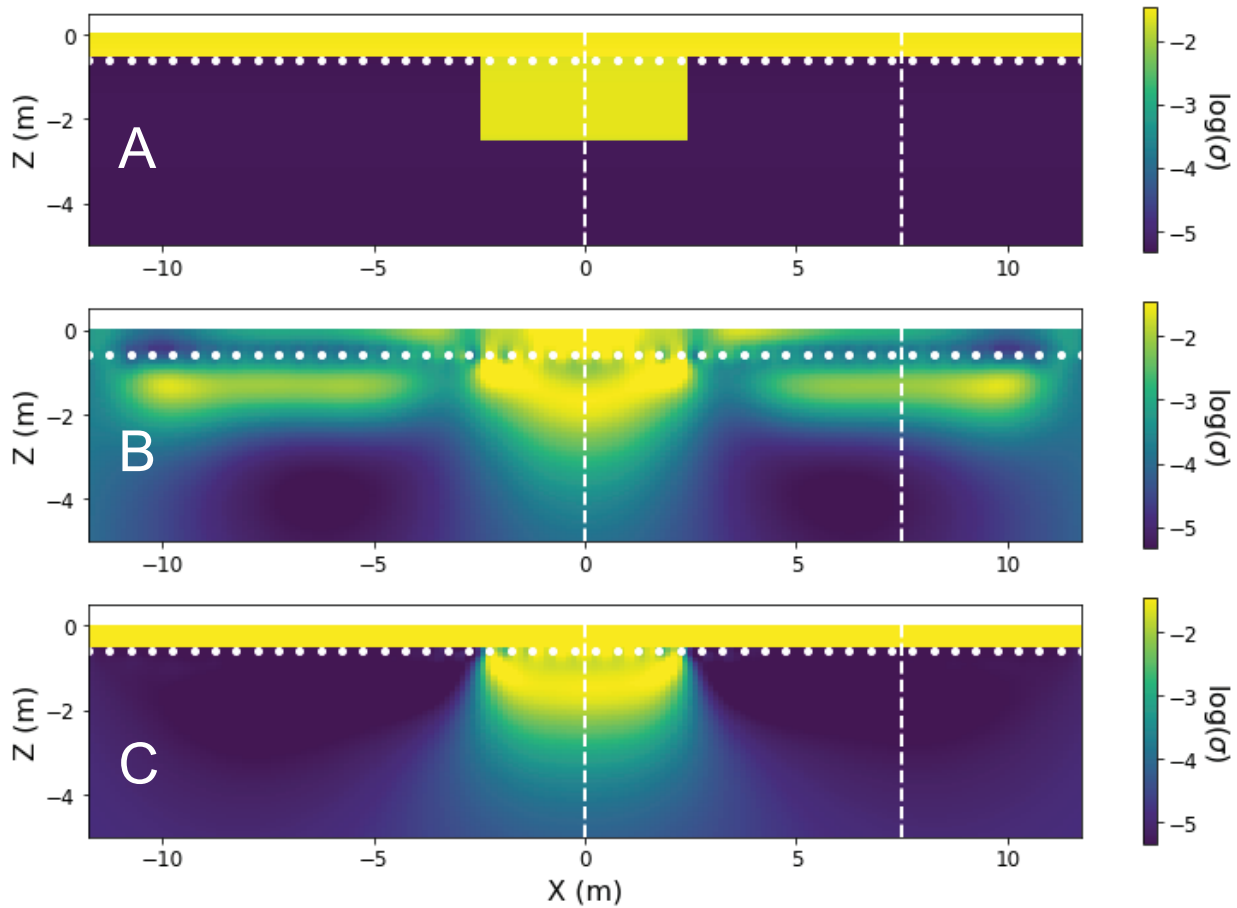

*Figure 9. Scenario 2 inversion results. A) true conductivity distribution, B) estimated conductivity model from a standard inversion, and C) estimated conductivity from a hybrid inversion. The standard inversion produced data with a 10% mean absolute percent error, while the hybrid inversion achieved a mean absolute data error of 1.1%. White dotted lines are the locations of the vertical profiles shown in Figure 10.*

Figure 10 shows vertical slices through the estimated models from the center of the array at $x$=0 m and off center at $x$=7.5 m. The difference between the standard and hybrid inversion is most apparent in the profiles outside the area of infiltration (i.e., $x$=7.5 m). These off-center profiles show that the hybrid inversion came much closer to recovering the magnitude and location of the surface water layer. On average, the smooth inversion underestimated the true average conductivity of the water layer by 104%, with up to 54% underestimates and 220% overestimates in some parts of the model space. On average, the hybrid inversion overestimated the true conductivity of the water layer by only 3.5%.

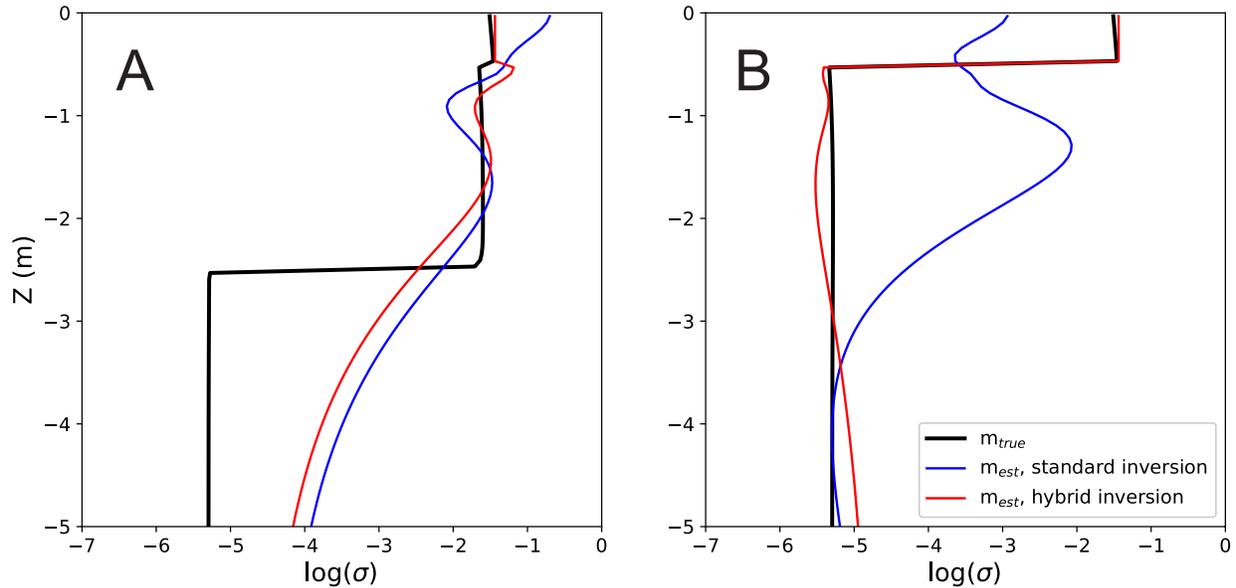

*Figure 10. Vertical profiles of the conductivity models shown in Figure 9 at locations of A) x=0 m and B) x=7.5 m.*

In summary, the hybrid inversion improves the estimated models for both scenarios presented here. The mean absolute errors for estimated models and associated data were calculated as follows:

$$err = \left| \frac{true - est}{true} \right| \times 100\% \qquad [3]$$

Table 2 shows that the mean absolute error of the estimated logarithmic conductivity in the region of interest (i.e., the *x* and *z* limits shown in Figure 5 and Figure 9) was much lower for the hybrid inversion than the standard inversion. Consequently, the misfit between the observed and predicted data was also much smaller in the hybrid inversion (Table 2).

*Table 2. Misfit of the estimated logarithmic conductivity model and corresponding data misfit for the standard and smooth inversions.*

|  | Scenario 1 | | Scenario 2 | |
|---|---|---|---|---|
|  | Standard | Hybrid | Standard | Hybrid |
| Mean absolute model error (%) | 20 | 3.7 | 33 | 10 |
| Mean absolute data error (%) | 10 | 2.1 | 9.6 | 1.1 |

# 6   Sensitivity Analysis

We performed a sensitivity analysis to evaluate how robust the hybrid inversion results were to a range of regularization parameters, initial models, and relative data errors. Sensitivity to regularization parameter $\beta$ was tested using a range of fixed values from $10^{-3}$ to $10^3$. Sensitivity to the starting model background ($m_{0\_bg}$) and upper layer ($m_{0\_layer}$) values was tested by perturbing initial values between -50 and 50%, where the initial value of the model background was the average apparent conductivity, and the initial layer value was the minimum apparent conductivity for scenario 1 and the maximum apparent conductivity for scenario 2. Sensitivity to data noise was tested using a range of values between 0 and 10%. Vertical profiles of the estimated conductivity models for each tested parameter value are shown in the supplementary material file.

To evaluate the sensitivity quantitatively, we calculated the mean absolute percent error in logarithmic electrical conductivity using Eq. [3] for each estimated model within the region of interest, i.e. the region shown in Figure 5 and Figure 9. The calculated errors shown in Figure 11 indicate the average sensitivity of the estimated model to each parameter. In both scenarios, large values of regularization parameter $\beta$ tended to cause higher average errors in the estimated model, reinforcing the importance of choosing an appropriate $\beta$ value. Both scenarios showed a relatively low sensitivity to initial model background conductivity. The hybrid inversions for scenario 1 tended to have a higher error when the initial model's layer conductivity was much too conductive, while the scenario 2 results were less sensitive to this parameter. The relative data error caused some variability in the estimated model for both scenarios. Notably, in both scenarios the hybrid inversion produced a model with lower error than the standard inversion for all of the tested parameter values.

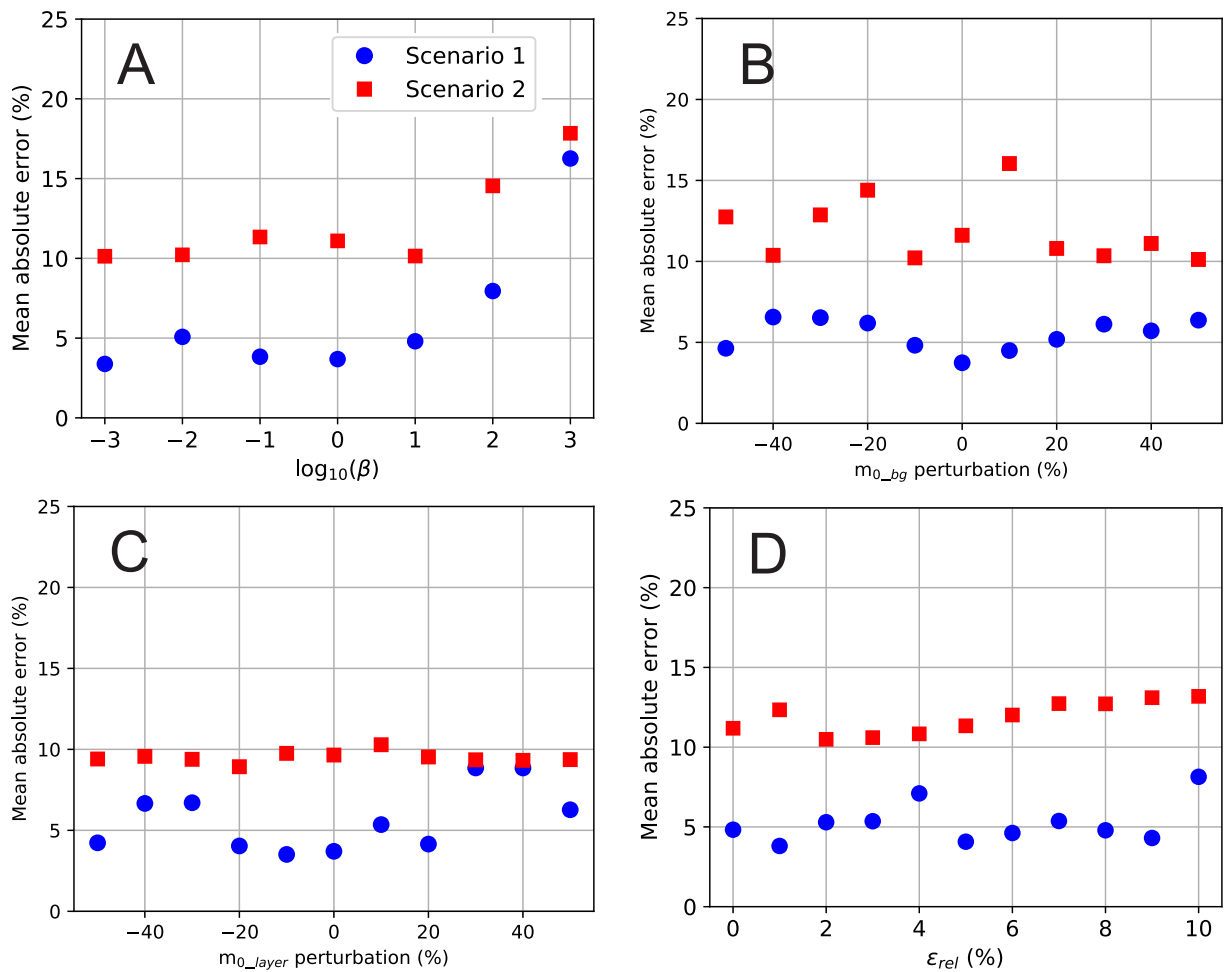

*Figure 11. Mean absolute percent error in logarithmic model conductivity in the region of interest for the range of tested regularization parameters (A), initial model background (B) and layer (C) values, and data errors (D). Values can be compared to the errors in the standard inversion, which were 20% for scenario 1 and 33% for scenario 2.*

# 7   Discussion

The hybrid inversion algorithm presented here is somewhat specific to the problem described, but the approach can be seen as a general framework that can be adapted to a variety of problems. For example, the layer-type hybrid parameterization used here could be applied to other scenarios, e.g., marine ERT where there is a water layer, permafrost surveys where there is a thawed active layer, or scenarios where electrodes are buried. The parameterization can also be modified to address different scenarios. If the depth to the interface is included in the inversion as a model parameter, the layer-type hybrid parameterization could also be used to estimate depth to interfaces, e.g., bedrock or the base of frozen ground. The hybrid methodology could be extended to other problems by using different parametric mapping functions to constrain parts of the model with shapes like blocks, circles, ellipsoids, etc. This could be useful for urban geophysical studies where the goal is to image the subsurface near tunnels or pipelines. However, the hybrid inversion may not be appropriate when subsurface features are not well constrained. If model parameterization uses incorrect assumptions about the geometry of subsurface features, the estimated model will be incorrect. In cases where the subsurface is not well constrained, a standard inversion is more appropriate.

Future work could also incorporate additional complexity in the parametric part of the model. In this study, a logical modification to the layer parameterization is to use a simple linear function to describe how conductivity changes as a function of depth. In this case, the layer's conductivity would be described by an intercept and slope instead of a single value. However, we found that this additional model complexity introduced nonuniqueness problems that made it difficult for the inversion to converge on an appropriate solution (Herring, 2021). While a detailed investigation is beyond the scope of this paper, future work will examine the degree of parameter independence to determine what level of complexity is justifiable in the parametric model. Future work could also explore how nonuniqueness problems could potentially be mitigated, e.g., by modifying the parametric model to incorporate constraints like upper and lower bounds or incorporating regularization on the parametric part of the model. Future work should also test the hybrid inversion algorithm on field data to further explore its advantages and potential pitfalls compared to standard data processing. Overall, the hybrid inversion approach presented here shows promise for improving the results of ERT surveys in a number of situations, and there are many opportunities for future research to test and improve this methodology.

## 8   Conclusions

The hybrid parametric/smooth inversion approach presented here improved electrical conductivity models by strategically incorporating prior structural information in the inversion. In this study, the model space was divided into a parametric layer at the surface and a smooth model below. Using this strategy, the hybrid inversion better resolved both highly resistive and conductive surface layers and the targets below, producing more accurate models with lower data misfit and fewer artifacts than the standard smooth inversion. The hybrid inversion produced similar resistivity models for a range of regularization parameters, starting models, and data noise levels, indicating fairly low sensitivity to these parameters. The hybrid inversion methodology is readily extensible to problems where well constrained parts of the subsurface can be parameterized with other simple geometric shapes, leading to a large range of potential applications.

# 9   Acknowledgements

This research was funded by a Mitacs Accelerate PhD fellowship [IT04279] in partnership with Matrix Solutions Inc., as well as the University of Calgary. We would also like to thank the associate editor and two anonymous reviewers for their helpful comments.

# 10  Data availability

The algorithm and data presented in this article can be found on Github at https://github.com/simpeg-research/Herring-2021-Hybrid-ERT-Inversion and have been archived in the Zenodo database at https://zenodo.org/record/5116927 (Herring et al. 2021).

# Supplementary Material

The hybrid inversion algorithm's sensitivity to regularization parameter $\beta$, initial model, and data noise was tested using a range of values for each parameter. The following figures show vertical profiles through the inverted models for each tested parameter value.

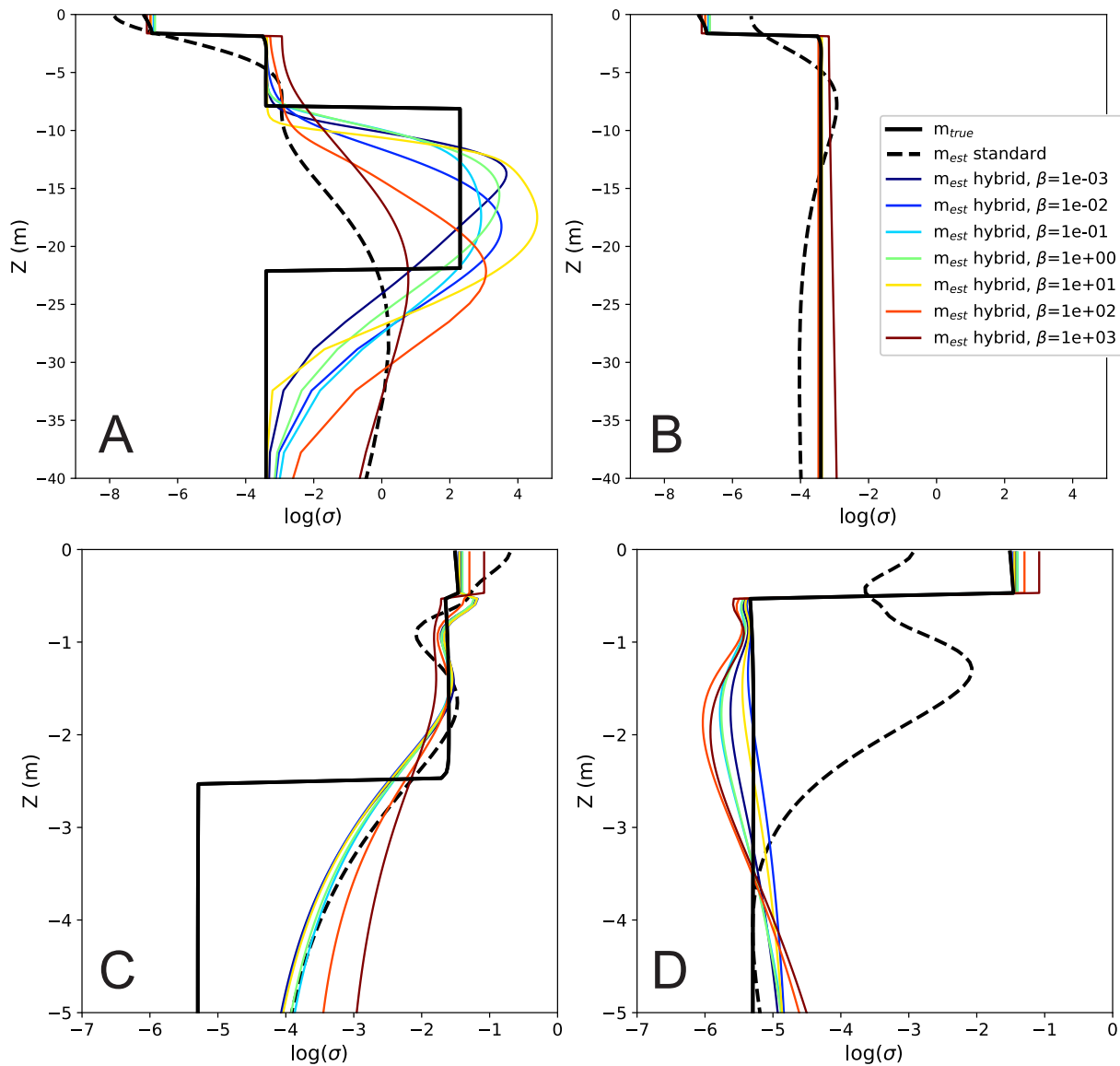

*Figure S1. Sensitivity of hybrid inversion results to regularization parameter $\beta$. A and B show vertical profiles for scenario 1 at x=0 m and x=75 m, respectively; C and D show vertical profiles for scenario 2 at x=0 m and x=7.5 m, respectively.*

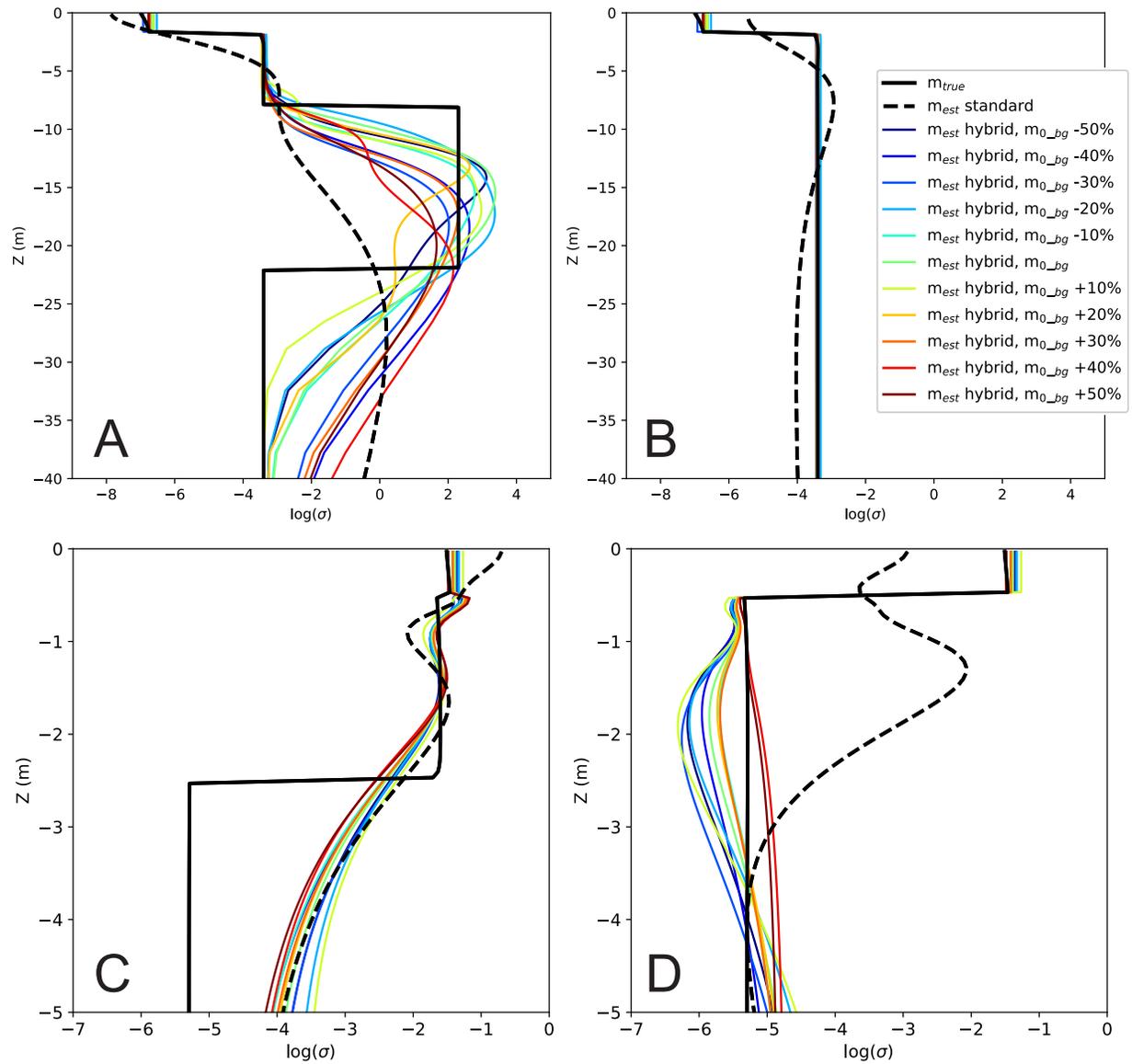

*Figure S2. Sensitivity of hybrid inversion to initial parametric model background conductivity value. The background conductivity was chosen to be the average apparent conductivity value of all data points, and sensitivity was tested by perturbing this value. A and B show vertical profiles for scenario 1 at x=0 m and x=75 m, respectively; C and D show vertical profiles for scenario 2 at x=0 m and x=7.5 m, respectively.*

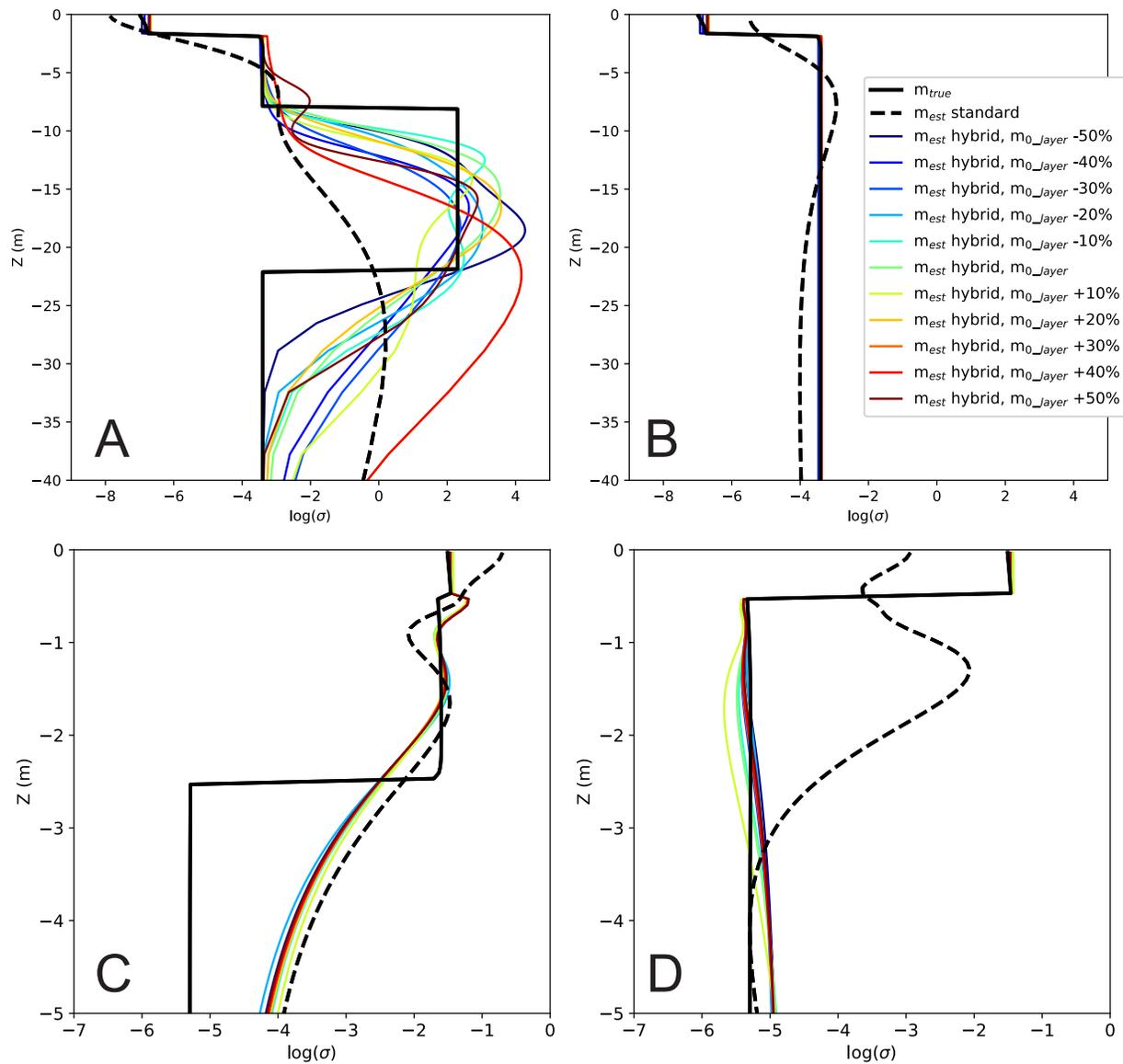

*Figure S3. Sensitivity of hybrid inversion to initial parametric model layer conductivity value. The layer conductivity was chosen to be the minimum apparent conductivity value of all data points for scenario 1 and the maximum apparent conductivity for scenario 2, and sensitivity was tested by perturbing this value. A and B show vertical profiles for scenario 1 at x=0 m and x=75 m, respectively; C and D show vertical profiles for scenario 2 at x=0 m and x=7.5 m, respectively.*

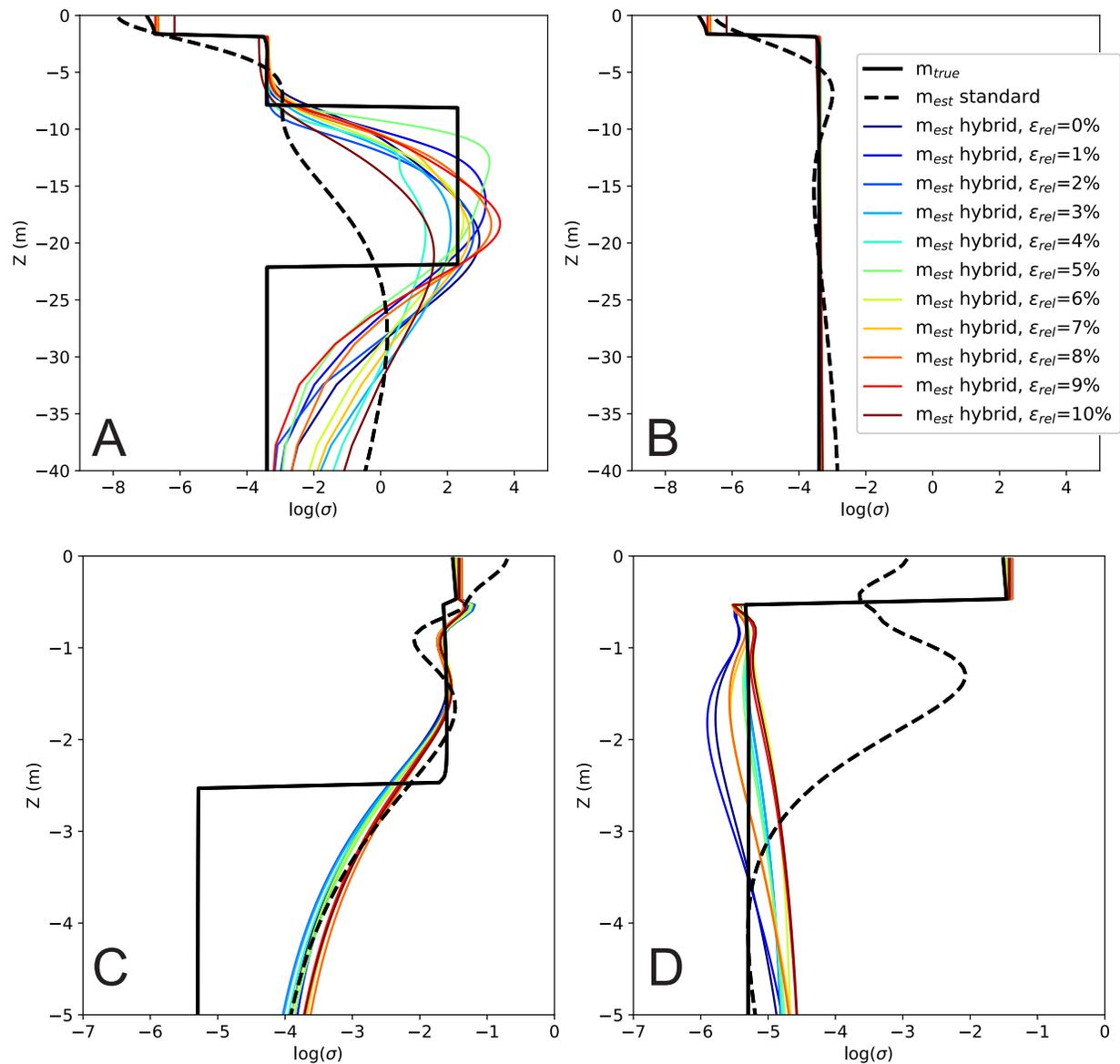

*Figure S4. Sensitivity of hybrid inversion results to relative data error. A and B show vertical profiles for scenario 1 at x=0 m and x=75 m, respectively; C and D show vertical profiles for scenario 2 at x=0 m and x=7.5 m, respectively. Note that the smooth inversion used data with a relative error of 1%.*